\documentclass[letterpaper, 10 pt, conference]{ieeeconf}  

\IEEEoverridecommandlockouts                              

\usepackage{graphicx} 
\usepackage{amsmath} 
\usepackage{amssymb}  
\usepackage{subfigure}
\usepackage{multirow}
\usepackage{todonotes}
\usepackage{caption}
\title{\LARGE \bf
Adversarially Trained Convolutional Neural Networks for Semantic Segmentation of Ischaemic Stroke Lesion using Multisequence Magnetic Resonance Imaging}

\author{Rachana Sathish$^{1}$, Ronnie Rajan$^{2}$, Anusha Vupputuri$^{1}$, Nirmalya Ghosh$^{1}$ and Debdoot Sheet$^{1}$
\thanks{$^{1}$R. Sathish, A. Vupputuri, N. Ghosh and D. Sheet are with Department of Electrical Engineering, Indian Institute of Technology Kharagpur, India - 721302
        {\tt\small rachana.sathish@iitkgp.ac.in}}%
\thanks{$^{2}$R. Rajan is with School of Medical Science and Technology, Indian Institute of Technology Kharagpur, India - 721302}%
}

\begin{document}

\maketitle
\thispagestyle{empty}
\pagestyle{empty}

\begin{abstract}

Ischaemic stroke is a medical condition caused by occlusion of blood supply to the brain tissue thus forming a lesion. A lesion is zoned into a core associated with irreversible necrosis typically located at the center of the lesion, while reversible hypoxic changes in the outer regions of the lesion are termed as the penumbra. Early estimation of core and penumbra in ischaemic stroke is crucial for timely intervention with thrombolytic therapy to reverse the damage and restore normalcy. Multisequence magnetic resonance imaging (MRI) is commonly employed for clinical diagnosis. However, a sequence singly has not been found to be sufficiently able to differentiate between core and penumbra, while a combination of sequences is required to determine the extent of the damage. The challenge, however, is that with an increase in the number of sequences, it cognitively taxes the clinician to discover symptomatic biomarkers in these images. In this paper, we present a data-driven fully automated method for estimation of core and penumbra in ischaemic lesions using diffusion-weighted imaging (DWI) and perfusion-weighted imaging (PWI) sequence maps of MRI. The method employs recent developments in convolutional neural networks (CNN) for semantic segmentation in medical images. In the absence of availability of a large amount of labeled data, the CNN is trained using an adversarial approach employing cross-entropy as a segmentation loss along with losses aggregated from three discriminators of which two employ relativistic visual Turing test. This method is experimentally validated on the ISLES-2015 dataset through three-fold cross-validation to obtain with an average Dice score of 0.82 and 0.73  for segmentation of penumbra and core respectively.

\end{abstract}

\section{Introduction}
\label{sec:intro}

Cerebrovascular accident (CVA), more commonly known as stroke, is one of the most common causes of death and disability in the world \cite{adamson2004stroke}. It is characterized by a sudden focal neurological deficit due to cerebral infarction caused by poor vascular supply. Ischaemic stroke is the most common type of stroke \cite{warlow1998epidemiology}, caused by occlusion of a blood vessel due to atherosclerotic stenosis, or by an embolus of atherosclerosis in a large artery, or may be of cardiac origin. This perfusion deficit causes irreversible necrosis of a small area of cerebral tissue which becomes completely devoid of blood supply. This area is called the core of the lesion and is surrounded by an area of hypoperfusion, which develops a reversible functional impairment due to temporary hypoxia. If the perfusion is not restored, this surrounding area which is called the penumbra of the lesion undergoes delayed apoptosis over the following days and weeks to form a permanent structural lesion with irreversible loss of function \cite{dirnagl1999pathobiology}. Early intervention to restore perfusion to this salvageable area can reverse the impairment and prevent the extension of the lesion \cite{atlantis2004association}. Therefore, one of the most crucial investigations into stroke lesion detection is evaluating the extent of penumbra as compared to the core of the lesion, which helps the physician decide on interventions like thrombolytic therapy. Multisequence magnetic resonance imaging (MRI) is classically employed, especially using perfusion-weighted imaging (PWI) and diffusion-weighted imaging (DWI), where PWI indicates the region of hypo-perfusion as in core and the penumbra, while DWI indicates the region of restricted diffusion as in core.

\begin{figure}[t]
\centering
\includegraphics[width=\columnwidth]{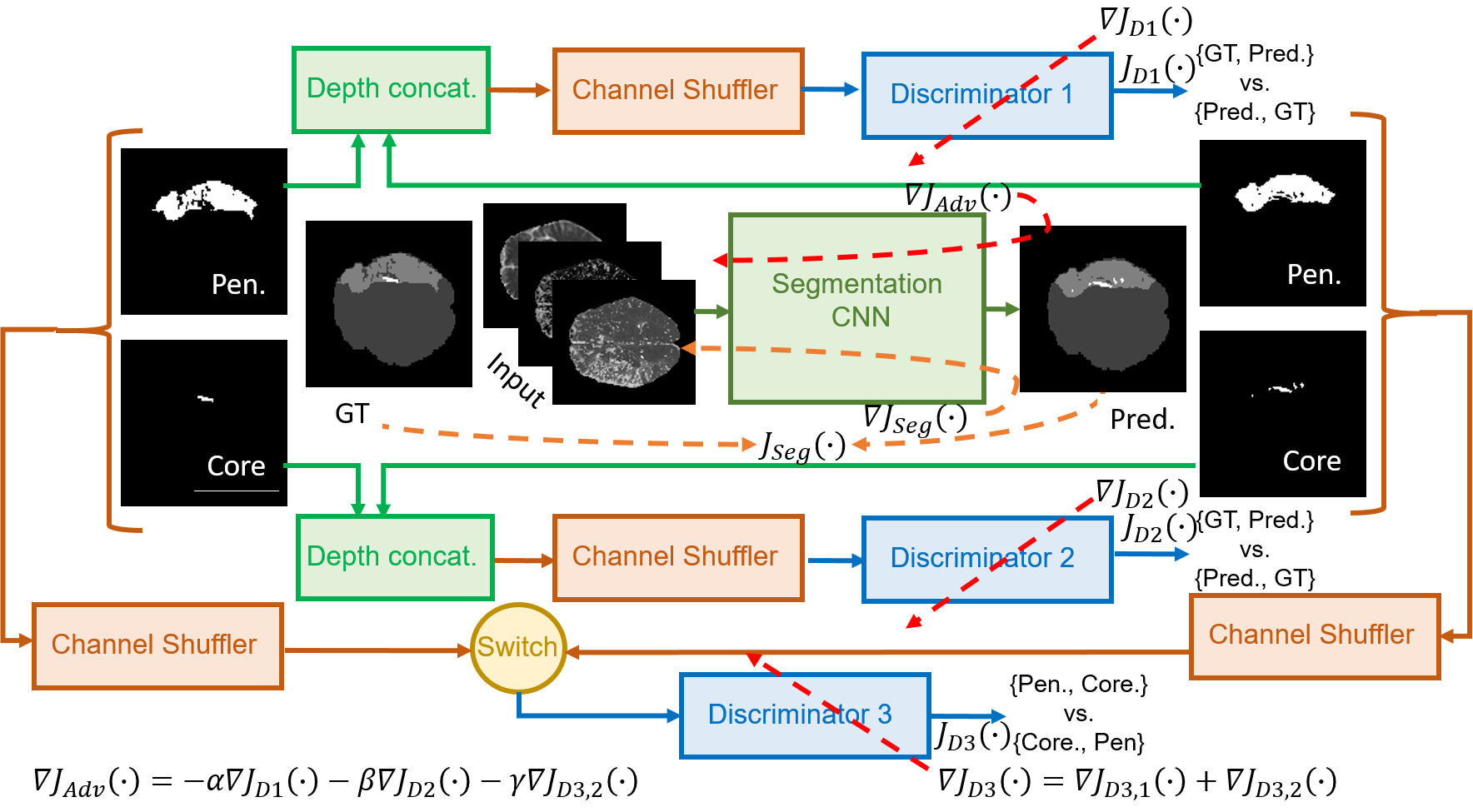}
\caption{Overview of the proposed method}
\label{fig:GA}
\end{figure}

\textbf{Challenge:} There is a wide range of variability in visual appearance across these sequences, which along with growth in the number of sequences makes symptomatic biomarker discovery for clinical use challenging.

\textbf{Approach:} In this context we propose a method for core and penumbra estimation employing recent developments in deep learning (DL) employing adversarial learning of a convolution neural network (CNN) for semantic segmentation as illustrated in~Fig.~\ref{fig:GA}. The limited availability of annotated data in stroke segmentation makes it difficult to train deep neural networks for automated detection with good generalisability and hence an adversarial approach is employed.

\begin{figure*}[t]
    \centering
    \subfigure[Phase 1: Train with segmentation loss]{\includegraphics[width=0.44\textwidth]{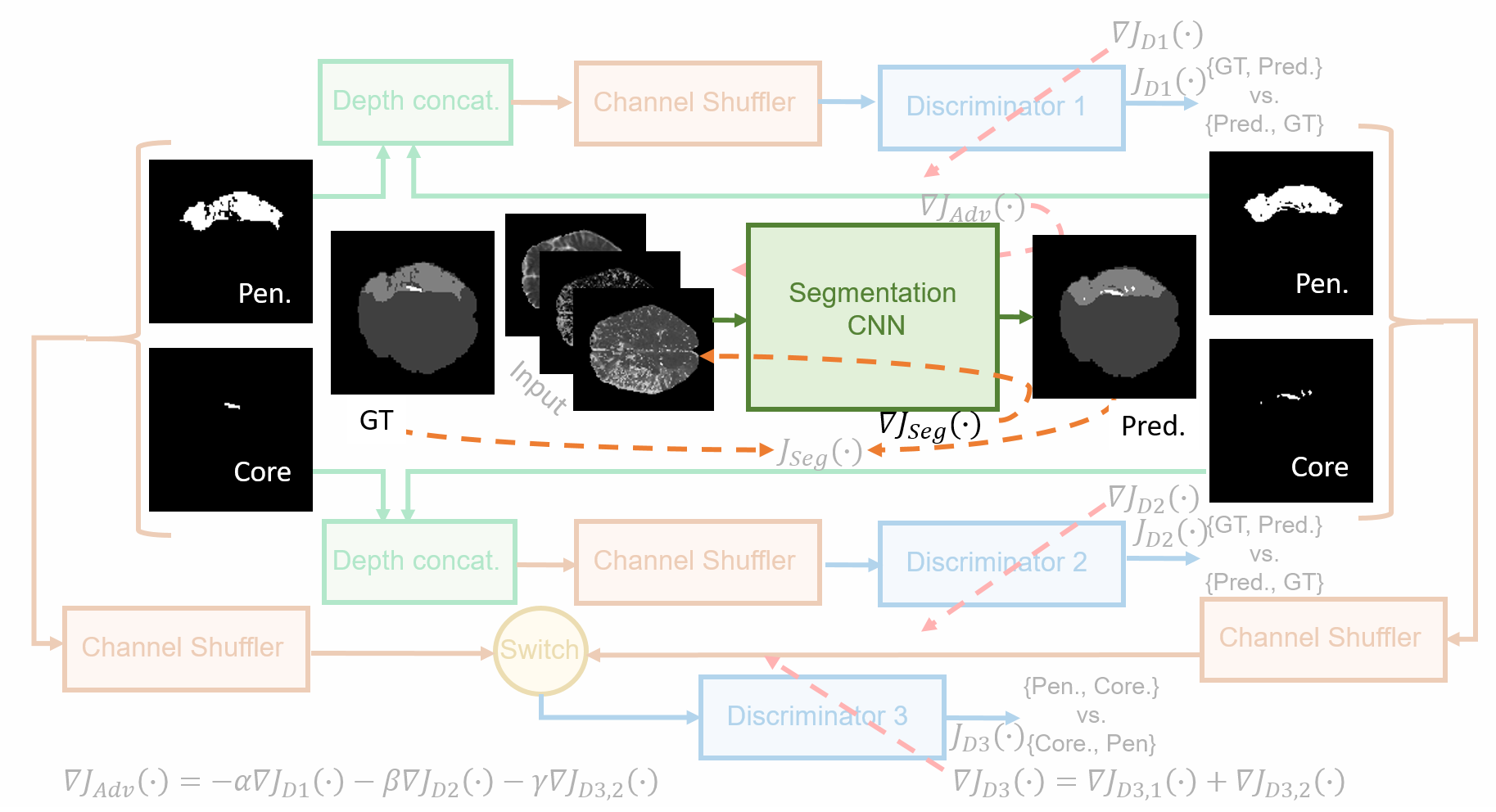}\label{p1}}
    \subfigure[Phase 2: Train discriminator 1]{\includegraphics[width=0.44\textwidth]{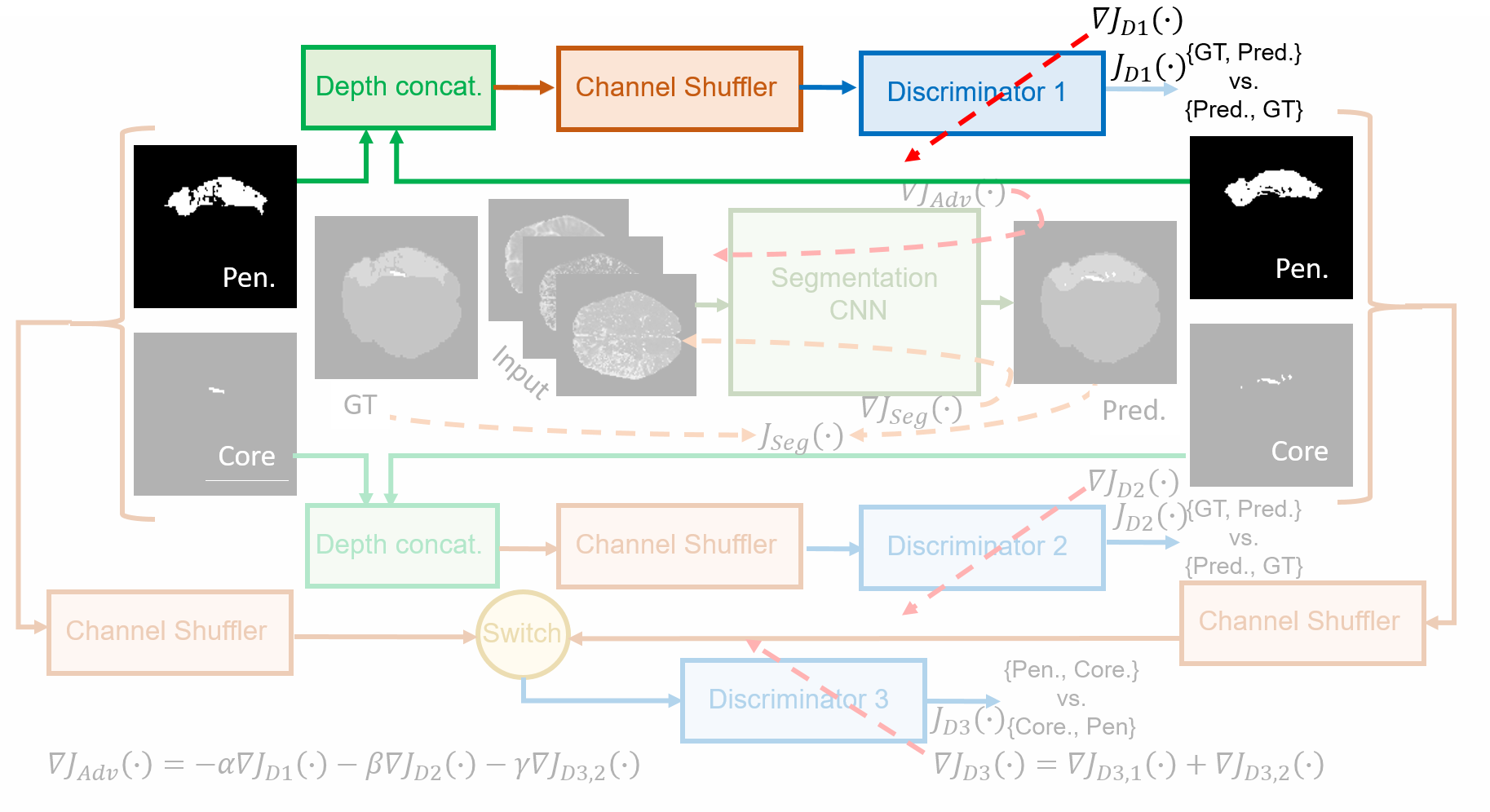}\label{p2}}

    \subfigure[Phase 3: Train discriminator 2]{\includegraphics[width=0.44\textwidth]{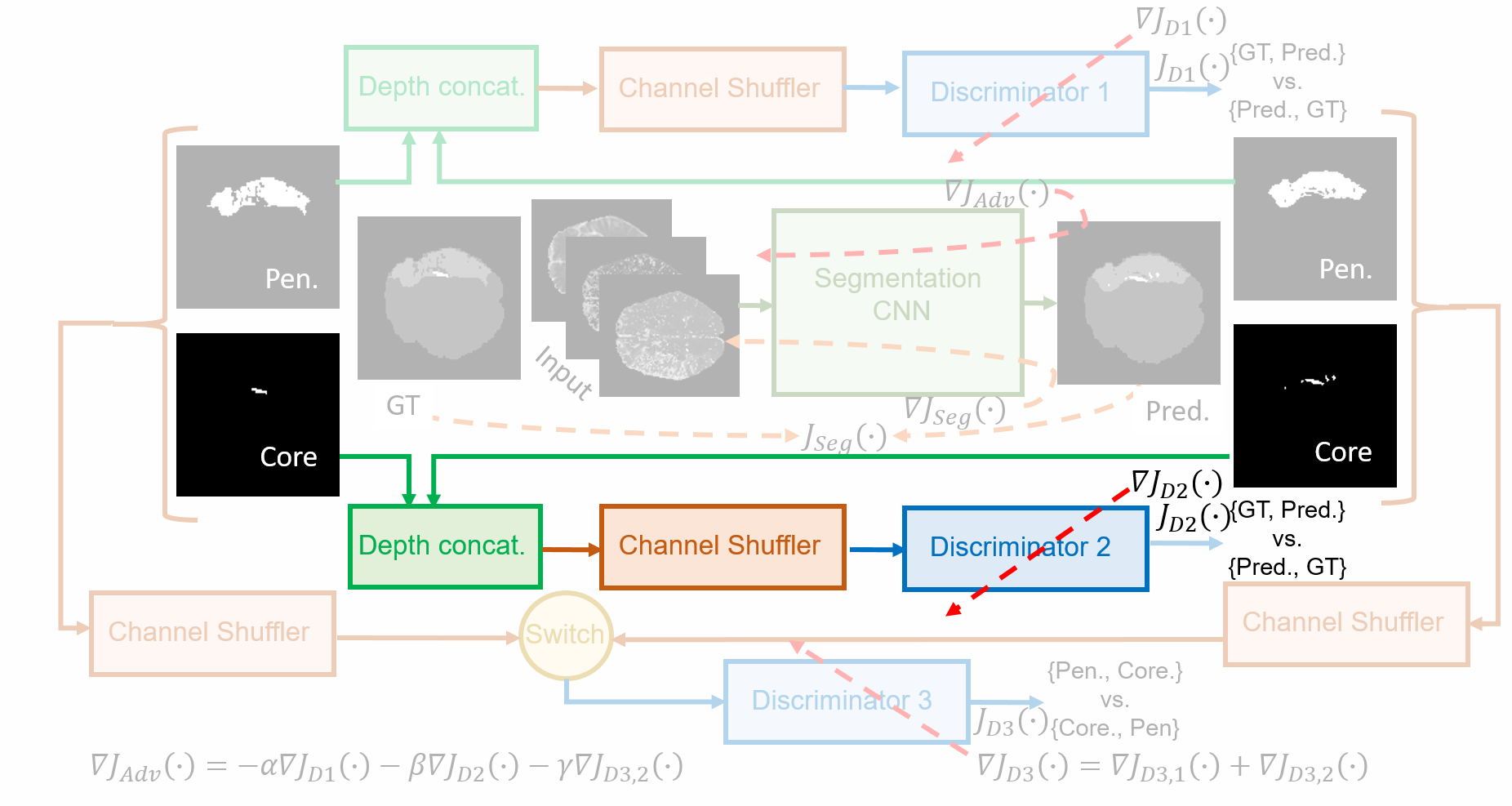}\label{p3}}
    \subfigure[Phase 4: Train discriminator 3]{\includegraphics[width=0.44\textwidth]{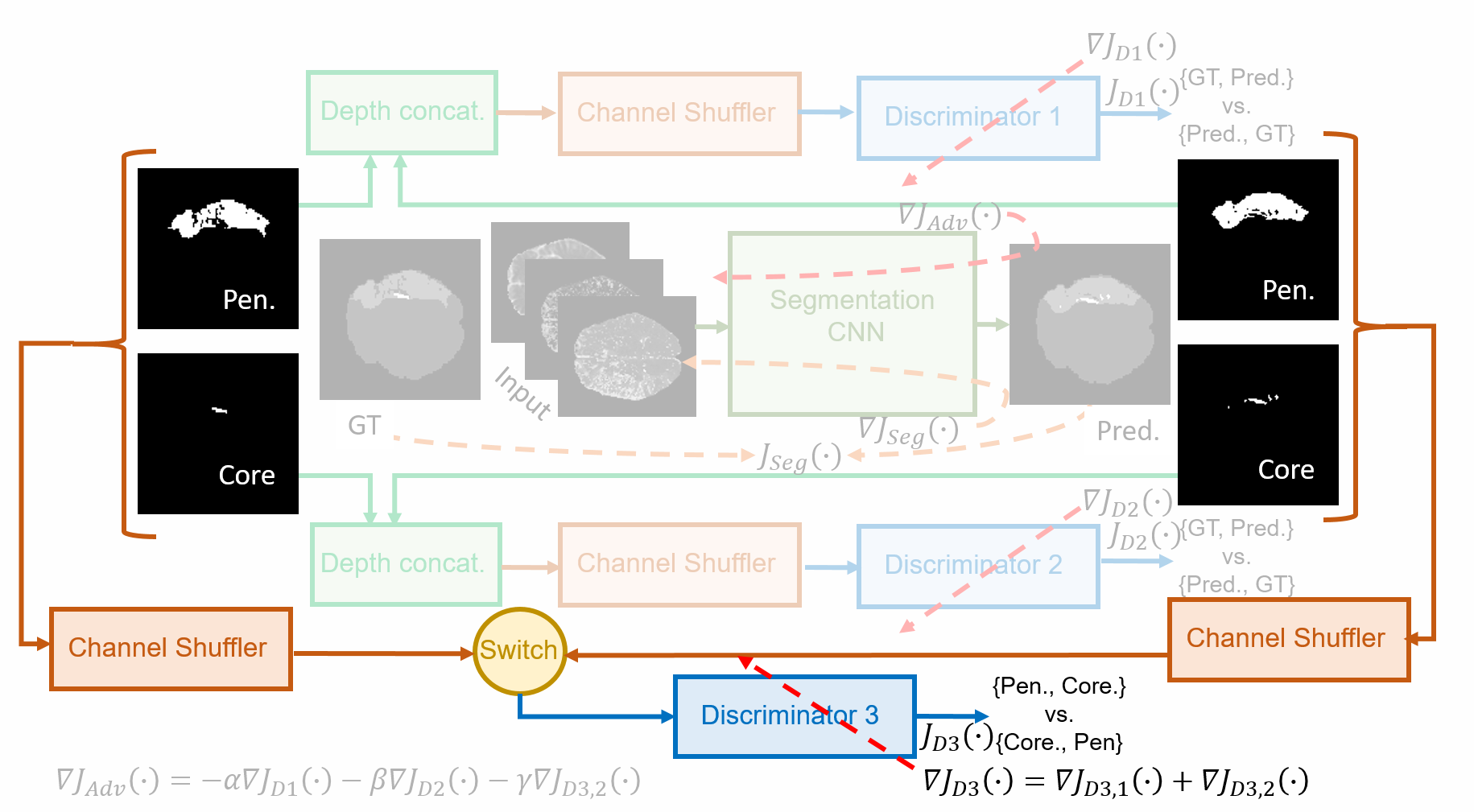}\label{p4}}
    
    \subfigure[Phase 5: Train with adversarial loss]{\includegraphics[width=0.44\textwidth]{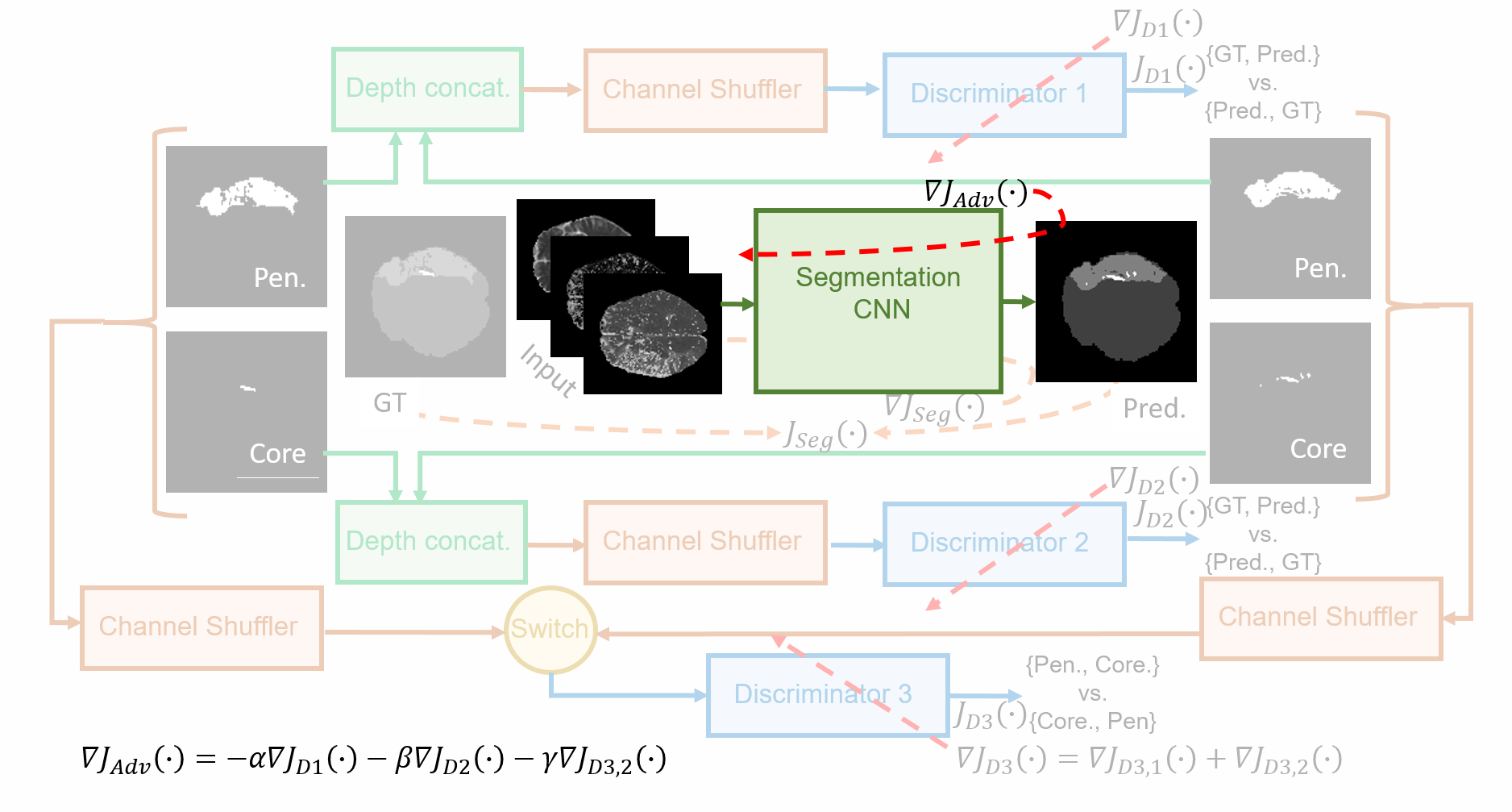}\label{p5}}
    \subfigure[Effect of including adversarial loss during training]{\includegraphics[width=0.44\textwidth]{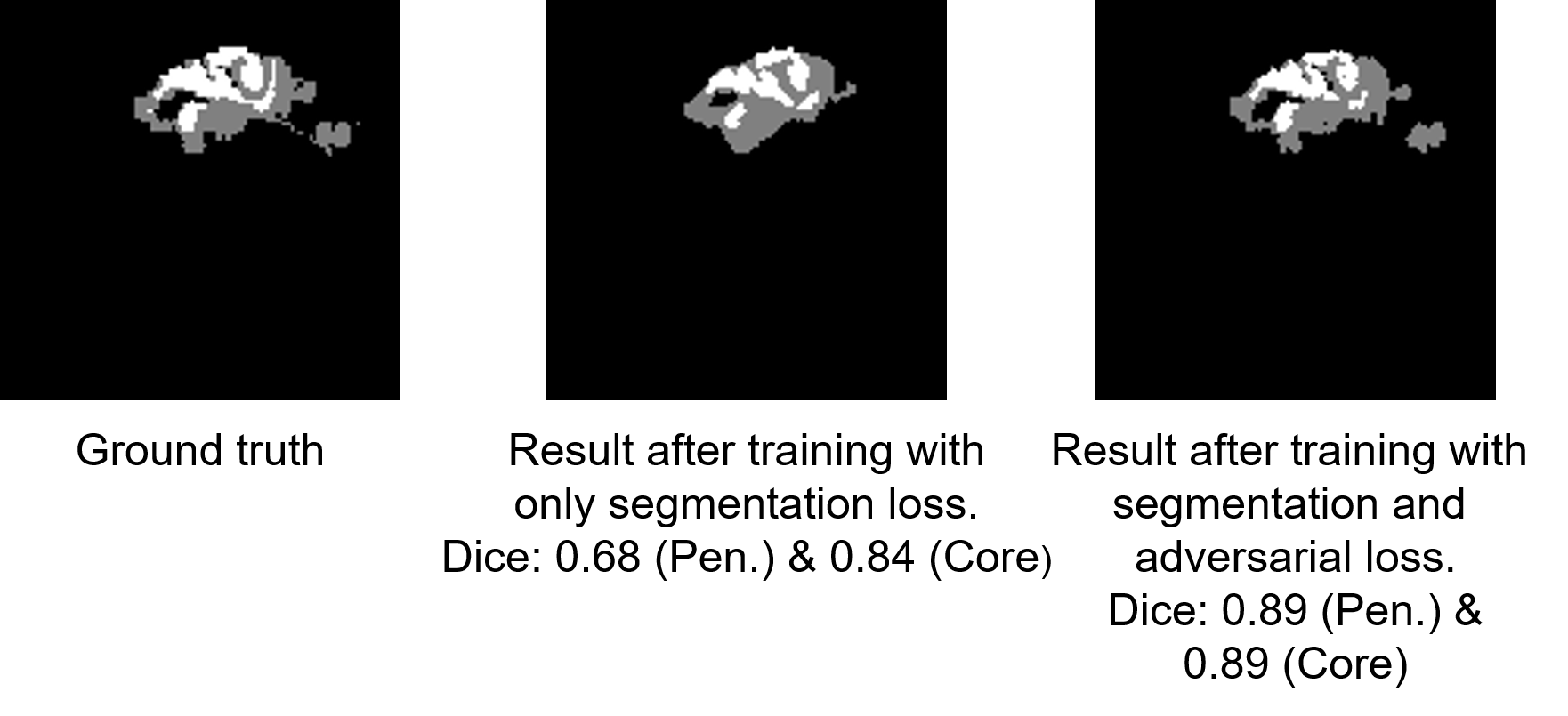}\label{adv_eff}}
    \caption{Training of the proposed framework using adversarial losses from three relativistic discriminators in addition to segmentation loss, for boosting ability to detect file lesions.}
    
    \label{fig:adv}
\end{figure*}


\textbf{Related work:} Random Fields (RF) based techniques have been most commonly employed for penumbra segmentation \cite{maier2017isles,halme2015isles,reza2015ischemic} on the ISLES-2015 dataset. Patch based stacked sparse auto-encoder for feature learning followed by a support vector machine (SVM) classifier have also been employed as a data-driven approach~\cite{praveen2018ischemic}. A multi-scale 11-layer deep 3D CNN with a 3D conditional random field (CRF) based post-processing was proposed by \cite{kamnitsas2017efficient}. More recently, several other approaches \cite{winzeck2018isles} have been proposed based on a modification of UNet~\cite{ronneberger2015u} for later versions of the ISLES challenge on stroke lesion segmentation.

\textbf{Organization:} The problem statement is defined in Sec.~\ref{sec:prob} with the proposed solution detailed in Sec.~\ref{sec:method}. Sec~\ref{sec:expt} outlines the various experiments conducted for validation of performance. Results and their discussion of the results are presented in Sec.~\ref{sec:disc} and Sec.~\ref{sec:conc} concludes the work.

\section{Problem Statement}
\label{sec:prob}
We consider the task of lesion segmentation to be carried out on a per slice basis. Given all the MRI sequences for a particular slice, it is arranged as a $T\times M\times N$ sized tensor $\mathbf{I}$ with $T$ denoting the number of MRI sequences and $M\times N$ being the spatial size of each slice. We model the segmentation problem as that of a three class semantic segmentation where each pixel is classified as belonging to either of the \{un-annotated brain tissue and background\} denoted as class 0, or \{penumbra\} denoted as class 1 or \{core\} denoted as class 2.

\begin{table*}[t]
\centering
\caption{Performance comparison with baselines}
\label{tab:base}
\begin{tabular}{|c|c|c|c|c|c|c|}
\hline
\multirow{2}{*}{Baseline} & \multicolumn{2}{c|}{Dice} & \multicolumn{2}{c|}{Precision} & \multicolumn{2}{c|}{Recall} \\ \cline{2-7} 
 & Pen. & Core & Pen. & Core & Pen. & Core \\ \hline
BL1 & $0.44 \pm 0.38$  & $0.10 \pm 0.17$ & $0.47 \pm 0.41$ & $0.14 \pm 0.12$ & $0.42 \pm 0.37$ & $0.15 \pm 0.26$ \\ \hline
BL2 & $0.77 \pm 0.04$ & $0.33 \pm 0.31$ & $0.75 \pm 0.06$ & $0.39 \pm 0.35$ & $0.80 \pm 0.05$ & $0.30 \pm 0.29$ \\ \hline
BL3 & $0.76 \pm 0.05$ & $0.47 \pm 0.41$ & $0.78 \pm 0.03$ & $0.75 \pm 0.22$ & $0.75 \pm 0.08$ & $0.50 \pm 0.44$ \\ \hline
Proposed & $0.82 \pm 0.06$ & $0.73 \pm 0.05$ & $0.82 \pm 0.05$ & $0.80 \pm 0.08$ & $0.83 \pm 0.08$ & $0.68 \pm 0.08$ \\ \hline
\end{tabular}
\end{table*}

\section{Method}
\label{sec:method}

In \textbf{Phase 1} the segmentation CNN ($\mathtt{net_{seg}}(\cdot)$) in Fig.~\ref{p1} predicts $\mathbf{\hat{O}}=\mathtt{net_{seg}}(\mathbf{I})$ where $\mathbf{\hat{O}}$ is a tensor of size $C\times M\times N$ with $C$ denoting the number of tissue classes, and the objective is to minimize the cross entropy loss $J_{Seg}(\cdot)$ between $\mathbf{\hat{O}}$ and $\mathbf{O}$, where $\mathbf{O}$ is the ground truth. Subsequently in \textbf{Phase 2} the first relativistic Turing test discriminator in Fig.~\ref{p2} learns to identify the ground truth (GT) annotation for penumbra from the segmented map obtained from ($\mathtt{net_{seg}}(\cdot)$) by minimizing the binary cross entropy loss $J_{D1}(\cdot)$. Similarly in \textbf{Phase 3} the second discriminator in Fig.~\ref{p3} learns to identify GT annotation of core from the segmented map by minimizing the binary cross entropy loss $J_{D2}(\cdot)$. In \textbf{Phase 4} the discriminator learns to predict which channel contains the penumbra when fed with a shuffled channels in the input as in Fig.~\ref{p4} thus minimizing the binary cross entropy loss $J_{D3}(\cdot)$. Finally in \textbf{Phase 5} the ($\mathtt{net_{seg}}(\cdot)$) parameters are optimized to be minimize the adversarial loss $J_{Adv}(\cdot)=-\alpha J_{D1}(\cdot)-\beta J_{D2}(\cdot)-\gamma J_{D3}(\cdot)$ and thus learn to be able to produce segmentation which closely resembles the GT. This impact of incorporating the adversarial losses in learning is finely visible in Fig.~\ref{adv_eff} where finer details of core and penumbra are evident in our segmentation approach.

\textbf{Segmentation CNN:} The segmentation CNN ($\mathtt{net_{seg}}(\cdot)$) used is an encoder-decoder like architecture~\cite{nandamuri2019sumnet} with the encoder having layer definitions similar to that of VGG11 \cite{Simonyan15}. Concatenation of features across matched layers in the encoder and decoder is present in this architecture along with the passing of max pooling indices for up-sampling in the decoder. We additionally add batch normalization after each convolutional layer. The VGG11 like encoder is initialized with ImageNet pre-trained model weights. 

\textbf{Discriminator networks:} The three discriminators ($D1, D2, D3$) are a shallow convolutional neural network with five convolutional layers each with $4\times 4$ kernels, interleaved with batch normalization layers and leaky ReLU non-linearity. Sigmoid activation is added to the last layer. The first layer has 32 channels. The number of channels in the subsequent layers is multiplied by a factor of 2.

\section{Experiments}
\label{sec:expt}

\textbf{Dataset description:} This method is experimentally validated using the Ischaemic Stroke Lesion Segmentation Challenge (ISLES) - 2015 dataset\footnote{www.isles-challenge.org/ISLES2015/}. We have used the SPES dataset from the challenge which consists of data from 30 subjects with an average of 70 slices per patient. Seven sequence maps viz. T1c, T2, DWI, CBF, CBV, TTP and Tmax of $94 \times 110$ size on average is available for each patient. We have trained our network using only the DWI, TTP and Tmax. Whitening transform is performed on the slices using the mean and standard deviation of corresponding sequences in the training set. Performance evaluation was conducted using 3-fold cross-validation. In each fold, 20 subjects were used for training, 5 for validation and 5 for testing. 

\textbf{Baselines:} The following baselines are used for comparison of performance. \textbf{BL1:} SegNet\cite{badrinarayanan2017segnet} trained using only segmentation loss. \textbf{BL2:} SegNet trained using adversarial losses as employed in the proposed framework. \textbf{BL3:} SUMNet trained using segmentation loss only. The performance of the proposed method in comparison with the baselines is tabulated in Tab.~\ref{tab:base}. The reported scores are mean and standard deviation across the three folds on the held-out test set. The performance of the proposed method is evaluated in comparison with the baselines using Dice coefficient, precision, and recall.

\begin{figure*}
    \centering
    \subfigure[Tmax]{\includegraphics[width=0.18\textwidth]{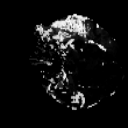}\label{tmax}}
    \subfigure[TTP]{\includegraphics[width=0.18\textwidth]{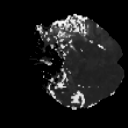}\label{ttp}}
    \subfigure[DWI]{\includegraphics[width=0.18\textwidth]{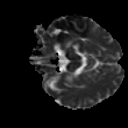}\label{dwi}}
    \subfigure[GT]{\includegraphics[width=0.18\textwidth]{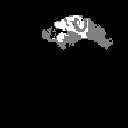}\label{gt}}
    
    \subfigure[SegNet Core - BL1]{\includegraphics[width=0.18\textwidth]{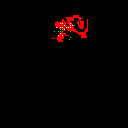}\label{bl3c}}
    \subfigure[SegNet Pen. - BL1]{\includegraphics[width=0.18\textwidth]{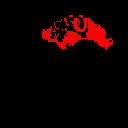}\label{bl3p}}
    \subfigure[SegNet Core - BL2]{\includegraphics[width=0.18\textwidth]{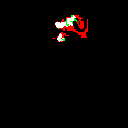}\label{bl4c}}
    \subfigure[SegNet Pen. - BL2]{\includegraphics[width=0.18\textwidth]{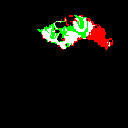}\label{bl4p}}
    
    \subfigure[SUMNet Core - BL3]{\includegraphics[width=0.18\textwidth]{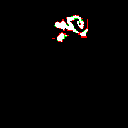}\label{bl1c}}
    \subfigure[SUMNet Pen. - BL3]{\includegraphics[width=0.18\textwidth]{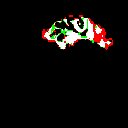}\label{bl1p}}
    \subfigure[Proposed Core]{\includegraphics[width=0.18\textwidth]{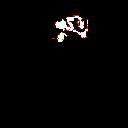}\label{bl2c}}
    \subfigure[Proposed Pen.]{\includegraphics[width=0.18\textwidth]{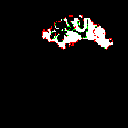}\label{bl2p}}
    
    \caption{Performance comparison for different baselines and our approach for estimating the core and penumbra using multisequence MRI. (a-e) denotes the different sequences used which constitutes the input to the network, (d) represents the GT with black representing the unannotated class 0, gray representing penumbra class 1 and white representing core as class 2. (e-f) represent the results of BL1, (g-h) for BL2, (i-j) for BL3 and (k-l) for our proposed approach. \textsc{Red} denotes under-segmentation, \textsc{Green} denotes over-segmentation and \textsc{White} denotes proper segmentation.}
    
    \label{fig:results}
\end{figure*}

\textbf{Training setup:} The segmentation network and the discriminators were trained using Adam optimizer for $200$ epochs, with a learning rate of $0.001$ without any augmentation of the dataset. In the adversarial loss, $\alpha=0.001$, $\beta=0.001$ and $\gamma=0.001$.

\section{Results and Discussion}
\label{sec:disc}
Qualitative results for the proposed method on a sample slice from the test set is illustrated in Fig.~\ref{fig:results} along with the sequences and the ground truth. The models compared were trained on the same fold of the data. It can be observed that the performance of segmentation improves significantly with the proposed adversarial training framework. There is a notable reduction in over-segmentation and under-segmentation.

Also, it can be seen from Tab.~\ref{tab:base} that the proposed method exhibits the least standard deviation in terms of dice coefficient across different folds of the dataset. This shows the generalization capability of the proposed framework despite being trained with a few annotated samples.

\section{Conclusion}
\label{sec:conc}
Early estimation of the extent of penumbra is one of the most crucial aspects of stroke management. This delineation between core and penumbra helps the physician decide on thrombolytic therapy that could reverse the damage to the salvageable tissue. Traditional methods of acute stroke lesion estimation utilize distinct image processing algorithms and handcrafted machine learning-based feature extraction techniques, to separately segment core and penumbra from DWI and PWI respectively. Our proposed method gives a unified framework that uses both diffusion and perfusion maps as inputs for deep learning based supervised learning of features to segment both core and penumbra with comparable accuracy. The limited availability of annotated diffusion and perfusion maps for the same patient has lead to over-fitting of networks in training data. This is mitigated by the use of adversarial learning.

\bibliographystyle{IEEEtran}
\bibliography{root}

\end{document}